\documentclass[useAMS,usenatbib]{mn2e}

\usepackage{graphicx,amssymb,natbib,cite,aas_macros,amsmath}		
\usepackage{graphics,times}

\def\etal{et al}	
\def\swift{\textit{Swift}}

\title[AMI Observations of GRB 130427A]{Probing the Bright Radio Flare and Afterglow of GRB 130427A with the Arcminute Microkelvin Imager}
\author[G. E. Anderson et al.] 
	{G.~E.~Anderson,$^{1,}$$^2$\thanks{E-mail: gemma.anderson@astro.ox.ac.uk} A. J. van der Horst,$^3$ T. D. Staley,$^{1,}$$^2$ R. P. Fender,$^{1,}$$^2$ R. A. M. J. Wijers$^3$
\newauthor	
A. M. M. Scaife,$^1$ C. Rumsey,$^4$ D. J. Titterington$^4$ A. Rowlinson,$^3$ R. D. E. Saunders$^4$\\
$^1$School of Physics \& Astronomy, University of Southampton, Southampton, SO17 1BJ\\
$^2$Department of Physics, Astrophysics, University of Oxford, Denys Wilkinson Building, Oxford, OX1 3RH\\
$^3$Astronomical Institute Anton Pannekoek, Science Park 904, P. O. Box 94249, 1090 GE Amsterdam\\
$^4$Astrophysics Group, Cavendish Laboratory, 19 J J Thomson Avenue, Cambridge CB3 0HE, UK}

\begin{document}

\date{Accepted 2014 March 6. Received 2014 March 5; in original form 2014 February 4}

\pagerange{\pageref{firstpage}--\pageref{lastpage}} \pubyear{2013}

\maketitle

\label{firstpage}

\begin{abstract}

We present one of the best sampled early time light curves of a gamma-ray burst (GRB) at radio wavelengths. Using the Arcminute Mircrokelvin Imager (AMI) we observed GRB 130427A at the central frequency of 15.7 GHz between 0.36 and 59.32 days post-burst. These results yield one of the earliest radio detections of a GRB and demonstrate a clear rise in flux less than one day after the $\gamma$-ray trigger followed by a rapid decline. This early time radio emission probably originates in the GRB reverse shock so our AMI light curve reveals the first ever confirmed detection of a reverse shock peak in the radio domain. At later times (about $3.2$ days post-burst) the rate of decline decreases, indicating that the forward shock component has begun to dominate the light-curve. Comparisons of the AMI light curve with modelling conducted by Perley et al. show that the most likely explanation of the early time 15.7 GHz peak is caused by the self-absorption turn-over frequency, rather than the peak frequency, of the reverse shock moving through the observing bands.

\end{abstract}


\section{Introduction}

The detection of the early time multi-wavelength radiation from gamma-ray bursts (GRBs) within the first day after the initial flash of $\gamma$-rays is essential for refining our understanding of these energetic events. The internal-external shock scenario \citep{piran99} suggests that along with the forward shock, which propagates into the circumburst medium to generate the classical afterglow, there is also emission associated with the reverse shock propagating into the relativistic ejecta \citep{sari99}. Evidence for the presence of a reverse shock has been demonstrated by the detection of optical flashes (within minutes after the $\gamma$-ray trigger) that are not correlated with the initial $\gamma$-ray emission from the GRB  \citep{sari99grb}. Such emission can only be explained by the presence of different physical emitting regions. This same model suggests that the detection of radio flares approximately 1 day post-burst also emanate from the reverse shock \citep{kulkarni99}. Such early time radio signals, which imply a rapid rise and fall in emission within 1 day post-burst, are atypical when compared to the classical radio afterglow of long GRBs resulting from the forward shock, which slowly evolve on the time-scales of days to years \citep[for a review see][]{granot14}.

The early time radio signature of GRBs has not been as well investigated as it has in the optical band. This is due to the limited number of large radio telescopes, which are required for such follow-up observations due to the faintness of GRB radio emission. This in turn makes it more difficult to acquire target-of-opportunity observations at the time of the event. Early time observations have traditionally required human intervention to activate, potentially resulting in the first radio observation of a given GRB being delayed several hours to even days post burst. As a result the radio emission emanating from the reverse shock of a GRB has only been observed in a few cases where the earliest radio detections have occurred around $1$ day post-burst \citep[e.g.,][]{kulkarni99,frail00,berger03}. 

Only a few robotized, rapid response, follow-up programmes of GRBs have been implemented in the radio domain. For example attempts were made with the Cambridge Low Frequency Synthesis Telescope at 151 MHz, which triggered on GRBs detected with the Burst And Transient Source Experiment onboard the \textit{Compton Gamma-Ray Observatory} \citep{green95,koranyi95,dessenne96}. More recently \citet{bannister12} conducted a robotized follow-up experiment using a 12~m radio dish at 1.4~GHz that was specifically designed to search for prompt radio emission associated with GRBs. This telescope triggered on those GRBs detected with the \swift\ $\gamma$-ray Burst Mission \citep{gehrels04} and was capable of being on target within a few minutes post-burst. In two out of the nine GRBs observed, a single dispersed radio pulse was possibly detected. In both cases the candidate's pulse was coincident with breaks in the GRB X-ray light curves. 

Over the past two years a new robotized follow-up programme using the Large Array (LA) interferometer of the Arcminute Microkelvin Imager \citep[AMI;][]{zwart08} has been implemented to obtain immediate observations, and conduct radio monitoring, of \swift\ detected GRBs at 15.7~GHz. This rapid GRB follow-up programme conducted with AMI is fully automated and is activated when \swift\ triggers on an event, with response times as low as 5 minutes \citep{staley13}. This programme is therefore capable of statistically constraining the radio properties of many \swift\ detected GRBs (both long and short) within the first hour post-burst, which has never been done before.

One of the most recent radio bright long gamma-ray bursts is GRB 130427A, which was detected on 2013, April 27 by both the Gamma-ray Burst Monitor \citep[GBM;][]{meegan09} onboard the \textit{Fermi Gamma-ray Space Telescope} at 07:47:06.42 UT \citep{zhu13,vonkienlin13}, and the Burst Alert Telescope \citep[BAT;][]{barthelmy05} onboard the \swift\ GRB mission at 07:47:57 UT \citep{maselli13}. GRB 130427A is situated at a redshift of $0.340$ making it the closest high-luminosity (E$_{\gamma,\mathrm{iso}} \gtrsim 10^{54}$ erg) gamma-ray burst since GRB 030329 \citep{levan13gcn}. Such nearby high-energy events are very rare as $\sim80\%$ of \swift\ GRBs are located at $z > 1$ and low redshift GRBs are often under-energetic \citep[see][and references therein]{perley14}. 

The extreme brightness and very early detection of an optical counterpart spurred a rapid succession of multi-wavelength follow-up observations making GRB 130427A one of the best spectrally and temporally sampled GRBs to date \citep{ackermann14,kouveliotou13,laskar13,levan13,maselli14,perley14,Xu13}. Broadband modelling conducted by \citet{laskar13}, \citet{perley14}, and \citet{panaitescu13}, using multi-wavelength observations ranging from 1 GHz to 0.1 TeV conducted between 300s and 60 days post-burst, revealed that the emission from GRB 130427A is best described by synchrotron emission from the combination of a reverse and forward shock. Extremely early optical observations conducted with RAPTOR (RAPid Telescopes for Optical Response) also detected a peak in optical emission $<20$s post-burst \citep{vestrand14}. This optical flash was temporally coincident with GRB 130427A's prompt $\gamma$-ray emission but modelling by \citet{vestrand14} demonstrated that it is more likely generated by the GRB's reverse-shock. This reverse shock was also shown to dominate the radio and mm wavelength bands from the first hours to days post-burst \citep{laskar13,perley14,panaitescu13}.

In this paper we present AMI observations of the energetic GRB 130427A starting at 0.36 days post burst, yielding the first early time ($<1$ day) radio detection of a GRB in the \citet{staley13} GRB follow-up observing campaign. The observations and data analysis are described in Section 2 with the resulting AMI fluxes and light curve presented in Section 3. In Section 3 we also present a basic broken power law fit to the AMI light curve of GRB 130427A and discuss how it compares to other early time radio detections of GRBs. The AMI light curve modelling is further discussed in Section 4 where we consider the implications of the different slopes, the position of the peak, and how our results compare to the Very Large Array (VLA) light curves and modelling of GRB 130427A conducted by \citet{perley14}. In Section 5 we summarise our findings. 

\section{Observations and Data Analysis}

GRB 130427A was observed with the AMI LA as part of the robotized GRB follow-up programme described by \citet{staley13}. The effective frequency range of this telescope is $13.9-17.5$ GHz when using frequency channels $3-7$, each with a 0.72 GHz bandwidth (channels 1,2 and 8 are disregarded due to their susceptibility at present to radio interference). AMI LA measures a single linear polarisation ($I+Q$) and has a flux RMS noise sensitivity of $3.3$~mJy s$^{-1/2}$ for 5 frequency channels. The LA consists of eight 12.8~m dishes with baseline lengths between 18-110~m, yielding a primary beam and angular resolution of $5.5'$ and $\approx30"$, respectively, at 15.7 GHz \citep{zwart08}. AMI is ideal for the initial follow-up of \swift\ GRBs as its field-of-view fully encompasses the position error of the BAT instrument \citep[$1-4'$;][]{barthelmy05}. In the follow-up AMI GRB monitoring programme further observations are then centred on the enhanced \swift\ X-ray Telescope \citep[XRT;][]{burrows05} position of the GRB (if X-ray emission is detected), which has a few arcsecond positional accuracy.

AMI was robotically triggered to observe GRB 130427A immediately following the \swift\ trigger. However, as the source was below AMI's horizon, the observation was automatically scheduled for 15:50:35 UT on 2013 April 27, $0.36$ days post-burst based on the GBM trigger, when GRB 130427A had reached a sufficient elevation. This first observation was of $1$ hour duration and took place during wet weather conditions; it was also at the LA's Eastern limit, which resulted in the loss of data due to pointing errors and antenna shadowing. The phase calibrator J1134+2901 was used for this observation, which was found to have a flux of only $50.82 \pm 5.09$ mJy in the AMI frequency range. In order to determine how much flux attenuation this first AMI observation suffered due to the low observing altitude and poor weather conditions, we compared the phase calibrator fluxes to those measured by the AMI Small Array \citep[which can be used to calibrate the LA,][]{franzen11} and they agree within 10\%. This error is taken into consideration when making flux uncertainty estimates on this dataset.

A further 13 AMI observations of GRB 130427A were then manually scheduled between 0.64 and 59.32 days using the phase calibrator J1125+2610, which is $\sim1$ Jy at 15.7 GHz. These subsequent AMI observations ranged in duration between $1-4$ hrs. The first stage of the data reduction involved using the python script \texttt{drive-ami} \citep[the basis of which was originally described in][]{staley13}, which analyses all the raw data files belonging to GRB 130427A. This script then runs the most up to date general AMI reduction pipeline, which uses the AMI \texttt{REDUCE} software to automatically flag for interference, shadowing and hardware errors, perform Fourier transforms of the lag-delay data into frequency channels, and then applies phase and amplitude calibrations \citep{perrott13}. The flux calibration was conducted using short observations of 3C286, 3C48, and 3C147. An additional flagging step was also performed on those visibilities for which one (or both) of the involved antennae have ``rain gauge'' correction values $<60\%$ of the nominal antenna value. The rain gauge system monitors the system temperature of each antenna to correct for increased noise due to atmospheric disturbances \citep{zwart08}. The $60\%$ cutoff is a very stringent limit that we have applied to the data to specifically target flux attenuation due to rain. This step flagged visibilities in the first April 27 observation as well as in the April 28, May 10 and June 25 observations. 

The resulting $uv-$FITS files output by AMI \texttt{REDUCE} were then imported into the Common Astronomy Software Applications package \citep[CASA:][]{jaeger08} for further analysis. Further interactive flagging of RFI missed by the AMI \texttt{REDUCE} pipeline was performed manually using the CASA task \texttt{plotms}. All the individual visibility datasets of GRB 130427A were then imaged with the CASA task \texttt{clean} using standard reduction techniques. A single clean box was specified, surrounding the position of GRB 130427A, within which was the only source in the field with a signal-to-noise ratio $>3$. The clean threshold was set to $\approx2$ times the root-mean square (RMS) of the image using a `Briggs' weighting scheme with $Robust = 0.5$.\footnote{See http://www.aoc.nrao.edu/dissertations/dbriggs/} The resulting cleaned images were then imported into Miriad \citep{sault95} where the peak flux of the source was measured using the task \texttt{imfit}, specifying a point source model. The measured flux error was calculated as the quadratic sum of the image RMS and the $5\%$ systematic flux calibration error of AMI \citep{perrott13}. This 5\% calibration error is conservative as the telescope has been found to be considerably better than this value \citep[for example see][]{scaife08,hurley-walker09,franzen11}.We assume a more conservative flux calibration error of 10\% for the first AMI observation at 0.36 days due to the low observing altitude, poor weather conditions and the use of a faint phase calibrator (J1134+2901) for this observation. As an additional check we have measured the flux density of GRB130427A from each AMI observation, directly from the visibilities, using different levels of automatic and interactive flagging, and splitting individual observations in time. We find that the flux densities measured in this way are fully consistent with the values obtained by imaging. 

\section{Results}

\subsection{Identification of the GRB counterpart}

In each AMI observation, a single unresolved source was detected, with small deviations in position as expected from the system noise. The absolute position accuracy of the AMI LA has also been studied. For example in the Tenth Cambridge Survey of radio sources at 15.7 GHz \citep[10C;][]{franzen11}, which was conducted with the AMI LA using a ``rastering'' technique, it was determined that the sources detected with a signal-to-noise ratio (SNR) $>5$ have an approximate position error of $3-4"$ \citep{davies11}. In 7 out of the 10 AMI observations of GRB 130427A where the point source mentioned above had a SNR $>5$, its corresponding Miriad \texttt{imfit} position was $<4"$ from the European VLBI Network position of GRB 130427A \citep{paragi13}. The remaining three were within $6.6"$ of the VLBI position. For each of the 14 AMI observations we calculated the Right Ascension (RA) and Declination (Dec) RMS position error resulting from the Miriad \texttt{imfit} of this source using Equations (4a), (4b), and (5) from \citet{perrott13}. In all but one AMI observation the offset between the Miriad \texttt{imfit} and VLBI position was less than three times the quadratic sum of the RA and Dec RMS errors. However, as the Miriad \texttt{imfit} position of the remaining detection of the source is $<4"$ from the majority of the other AMI detections, it is extremely likely that it is the same source in all 14 AMI observations. We therefore identify this source as the radio afterglow of GRB 130427A.

The final results for each AMI observation of GRB 130427A are listed in Table~\ref{tab:1}. This table includes the observing time of each AMI observation, the number of days the observation took place after the \textit{Fermi} GBM trigger, and the flux measured with Miriad \texttt{imfit}. 

\begin{table}
\begin{center}
 \caption{AMI Observations of GRB 130427A}
 \label{tab:1}
 \begin{tabular}{ccc}
 \hline
Obs. time$^{\mathrm{a}}$ & Days since burst$^{\mathrm{b}}$ & Flux density$^{\mathrm{c}}$ \\
 (UT) & & (mJy) \\
\hline
2013/04/27.68  &     0.36  & $3.44 {~\pm~} 0.38$ \\
2013/04/27.96  &     0.64  & $4.16 {~\pm~} 0.22$ \\
2013/04/28.88  &     1.55  & $1.64 {~\pm~} 0.13$ \\
2013/04/29.80  &     2.48  & $1.03 {~\pm~} 0.09$ \\
2013/04/29.97  &     2.65  & $0.94 {~\pm~} 0.09$ \\
2013/04/30.88  &     3.55  & $0.61 {~\pm~} 0.06$ \\
2013/05/01.91  &     4.58  & $0.56 {~\pm~} 0.05$ \\
2013/05/02.85  &     5.53  & $0.51 {~\pm~} 0.06$ \\
2013/05/04.85  &     7.52  & $0.45 {~\pm~} 0.05$ \\
2013/05/07.97  &     10.64  & $0.35 {~\pm~} 0.05$ \\
2013/05/10.90  &     13.58  & $0.31 {~\pm~} 0.07$ \\
2013/05/21.77  &     24.44  & $0.28 {~\pm~} 0.05$ \\
2013/06/05.77  &     39.44  & $0.23 {~\pm~} 0.05$ \\
2013/06/25.65  &     59.32  & $0.12 {~\pm~} 0.04$ \\
\hline

\end{tabular}
\end{center}

$^{\mathrm{a}}$ The date of the observation corresponds to the mid point of the AMI integration.\\
$^{\mathrm{b}}$ The days since burst are relative to the initial \textit{Fermi} GMB detection of GRB 130427A.\\
$^{\mathrm{c}}$ The flux error is the quadratic sum of the image's rms and the $5\%$ systematic flux calibration error with the exception of the observation at 2013/04/27.68, which assumes a conservative $10\%$ calibration error (see Section 2).
\end{table}

\begin{table*}
\begin{center}
\caption{Brightness temperatures and minimum Lorentz factors calculated for GRBs with early time radio detections}
\label{tab:2}
  \begin{tabular}{cccccccc}
 \hline
GRB & Redshift & Luminosity & Frequency & Days since burst & Brightness & Minimum Lorentz & Reference$^{\mathrm{b}}$ \\
& & distance$^{\mathrm{a}}$ & & & temperature & factor & \\
& & (Gpc) & (GHz) & & ($\times10^{15}$K) & $\Gamma$ & \\
\hline
990123  & 1.6      & 11.7    &    8.46  &  1.24   &   1.9  & 12.3 & 1 \\
991216  & 1.02    & 6.7      &    15.0  &  1.33   &   0.9  & 9.7   & 2 \\ 
991216  & 1.02    & 6.7	 &    8.46  &  1.49   &   2.0  & 12.6 & 2 \\
010222  & 1.477  & 10.6    &    22.5  &  0.32   &   9.2  & 21.0 & 3 \\
010222  & 1.477  & 10.6    &    350.0 & 0.35   &   0.2  & 5.8   & 4 \\
020405  & 0.69    & 4.1      &    8.46  &  1.19   &   0.7  & 8.9   & 5 \\
130215A & 0.597 & 3.5      &    93.0  &  0.11   &   3.1  & 14.6 & 6 \\
130418A & 1.218 & 8.4	 &    93.0	&  0.34  &   1.4  & 11.2  & 7 \\	
130427A & 0.340 & 1.8      &    15.7 	&  0.36  &   3.7  & 15.4 & 8 \\
130427A & 0.340 & 1.8      &    5.1    &  0.68   &   3.6  & 15.4 & 9,10 \\
130907A & 1.238 & 8.5      &    24.5  &  0.17   &   33.2 & 32.1 & 11 \\
\hline

\end{tabular}
\end{center}

\begin{flushleft}

$^{\mathrm{a}}$ The luminosity distance was calculated from the redshift using the online cosmology calculator developed by \citet{wright06} assuming cosmological parameters $\rm{H}_{O}=72 \rm{~km~s}^{-1}\mathrm{Mpc}^{-1}$, $\Omega_{\rm{M}}=0.27$, and $\Omega_{\rm{vac}}=0.73$.\\
$^{\mathrm{b}}$ References for the radio detection used to calculate the brightness temperature for each GRB. 1: \citet{kulkarni99}; 2: \citet{frail00}; 3: \citet{frail03}; 4: \citet{frail02}; 5: \citet{berger03};  6: \citet{perley13gcn14210};  7: \citet{perley13gcn14387}; 8: This paper; 9: \citet{laskar13}; 10: \citet{perley14}; 11: \citet{corsi13gcn}

\end{flushleft}

\end{table*}

\subsection{Historical context and brightness temperature}

Historically the earliest radio detections of GRBs in the literature occur approximately 1 day post-burst. For example \citet{kulkarni99} detected a short lived radio flare from GRB 990123 peaking at $\sim1.24$ days post-burst. Rigorous radio observations both before and after this single detection only yielded upper-limits, which lead to \citet{kulkarni99} interpreting this brief radio counterpart appearance as emission from the GRB's reverse shock. Consequently similar observations were also conducted for GRB 991216 \citep{frail00} and GRB 020405 \citep{berger03}. In both cases early radio observations took place $\approx1$ day post-burst, resulting in the detection of a radio counterpart. These counterparts then rapidly faded but were still detectable several weeks post-burst. This rapid decrease in radio flux is atypical when compared to the rise and decay of most radio afterglows that result from the forward shock, which tend to peak between $3-100$ days post burst \citep{chandra12}. The reverse shock was therefore the most likely interpretation for the early time radio emission observed from GRB 991216 and GRB 020405 \citep[see][respectively]{frail00,berger03}.

The early time AMI detection of GRB 130427A at 0.36 days is one of the earliest detections of a long GRB at radio wavelengths (long GRBs that have earlier radio detections with significance $>3\sigma$ include GRB 010222 \citep{frail02}, GRB 130215A \citep{perley13gcn14210}, GRB 130418A \citep{perley13gcn14387}, and GRB 130907A \citep{corsi13gcn}). If we assume that the emission from GRB 130427A is a non-relativistic flow that emanates from a region of size $ct$ then a brightness temperature of $T_{b} = 3.7 \times 10^{15}$~K is calculated from the first AMI observation using
\begin{equation}  T_{\rm{b}} = 1.153 \times 10^{-8}\,d^{2}\,F_{\nu}\,\nu^{-2}\,t^{-2}\,(1+z)^{-1} \label{eq:1}
\end{equation}
where $d$ is the distance to the GRB in cm, $F_{\nu}$ the flux in Jy, $\nu$ is the observing frequency in Hz, $t$ is time in seconds since the $\gamma$-ray trigger, and $z$ the redshift \citep{longair11}. Since $ct$ is the maximum size of the emitting region, and the maximum brightness temperature is the inverse-Compton limit $T_{\rm{B}}\approx10^{12}$ K, then one derives a minimum Lorentz factor from the observed brightness temperature such that $T_{\rm{b}} / T_{\rm{B}} = \Gamma^{3}$ \citep{galama99}. The minimum bulk Lorentz factor predicted by the earliest AMI observation of GRB 130427A is therefore $\Gamma \gtrsim 15.4$, confirming its relativistic nature.

A comparison between the brightness temperature calculated from this first AMI observation of GRB 130427A and those calculated from the $\approx1$ day radio detections of the confirmed reverse-shock-detected GRBs (GRB 990123, GRB 991216, and GRB 020405), the four early time ($<0.36$ days post burst) radio detected GRBs mentioned above (GRB 010222, GRB 130215A, GRB 130418A, and GRB 130907A), and the earliest VLA detection of GRB 130427A, can be found in Table~\ref{tab:2}. This table lists the redshift of the GRB and its corresponding luminosity distance in Gpc calculated using \citet{wright06}, the frequency of the observation, the time of detection in days post-burst, the brightness temperature calculated using Equation~\ref{eq:1}, and the corresponding predicted minimum Lorentz factor. These results demonstrate that our early AMI observation of GRB 130427A represents one of the highest brightness temperatures and Lorentz factors based on radio observations. The AMI GRB robotic follow-up programme therefore has the potential to play an important role in increasing this sample significantly. 

\subsection{Description and fit to the light curve}

\begin{figure}
\begin{center}
\includegraphics[width=0.5\textwidth]{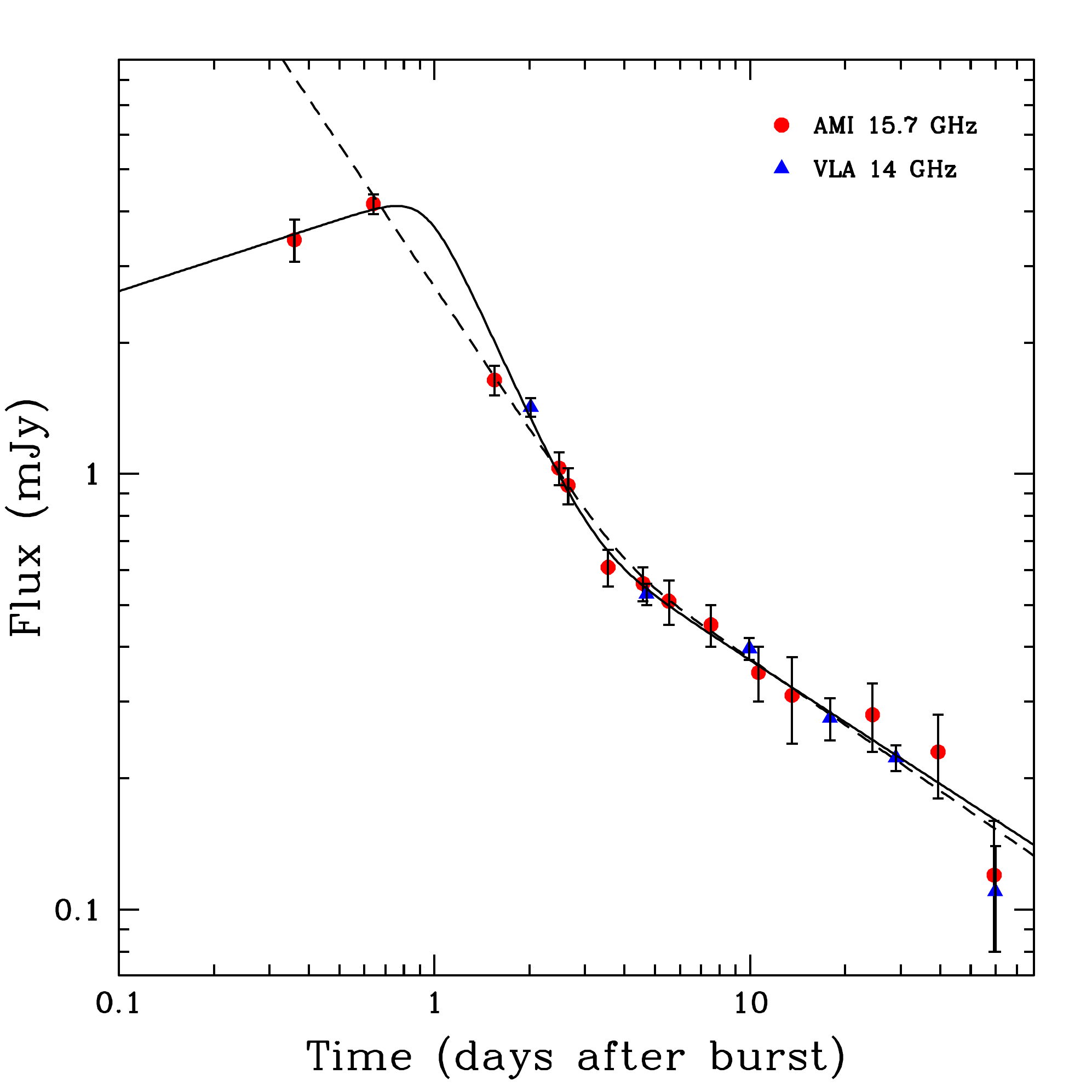}
\end{center}
\caption{The AMI 15.7 GHz light curve of GRB 130427A (red circles). Over-plotted are the VLA detections of GRB 130427A at 14 GHz (blue triangles). The solid line is the double broken power law fit to all the AMI and VLA data points. The dashed line is the single broken power law, which is a fit to all but the first data point. All errors are $1\sigma$.}
\label{fig:1}
\end{figure}

The peak in the AMI light curve of GRB 130427A around 1 day post-burst is a signature that has never been observed from a GRB at such early times in the radio frequency domain. Investigations by \citet{perley14} determined that the transition frequency between weak and strong scattering at the Galactic latitude of GRB 130427A is about 5 GHz, which places the AMI observing frequency in the weak scattering regime where the modulation index is only a few percent; scintillation is therefore unlikely to be the cause of the peak feature in the AMI light curve. Figure~\ref{fig:1} shows the AMI light curve of GRB 130427A (red circles), which is better sampled than the VLA $13-16$ GHz observations (blue triangles).

The afterglow from a GRB is generally well described by a decaying single or broken power law resulting from the relativistic ejecta decelerating as it sweeps up the ambient medium \citep{sari98,sari99jet}. In order to quantify the behaviour of the GRB 130427A AMI light curve, we fit two different broken power laws to the data. The initial broken power law fit ignores the first AMI data point, concentrating on the decaying part of the light curve. This fit results in a break time at 4.0 days with a pre-break temporal decay slope of $\alpha = -1.1$ and a post-break slope of $\alpha = -0.5$ ($\chi^{2}_{red} =1.09$) for $F(t) \propto t^{\alpha}$. This power law fit is shown as a dashed line in Figure~\ref{fig:1}. Clearly this broken power law does not describe the early time emission from GRB~130427A that we have observed with AMI as it predicts a 15.7 GHz flux of around $8$ mJy at 0.36 days post-burst. This fit overestimates the measured AMI flux at 0.36 days by a factor of 2.4, which is equivalent to $13$ standard deviations. As explained in Section 2 this is well outside the flux attenuation that we might expect to have occurred during this observation. We therefore confirm that the increase in 15.7 GHz flux $<0.64$ days post-burst  is a real feature.

We next fit a double broken power law to all the data points in the AMI light curve. This fit resulted in a temporal power law rise of $\alpha = 0.2 $,  peaking at 0.9 days, followed by two temporal decay slopes of $\alpha = -1.6 $ and $\alpha = -0.5 $ with a break time at 3.2 days ($\chi^{2}_{red} =1.06$). This double broken power law fit is depicted as a solid line in Figure~\ref{fig:1}. The earlier decay index is comparable to those calculated from power law fits to the early time ($<10$ days post-burst) radio detections of GRB 991216 \citep[$\alpha_{d} = -0.82 \pm 0.02$;][]{frail00} and GRB 020405 \citep[$\alpha_{d} = -1.2 \pm 0.4$;][]{berger03}. The AMI light curve of GRB 130427A is also clearly declining at $\sim1$ days post-burst, which is consistent with the radio light curves of GRB 990123, GRB 991216, and GRB 020405, and therefore with a reverse shock interpretation.

\section{Interpretation}

Broadband spectral modelling of GRB 130427A, which utilises observations from GHz radio wavelengths to GeV $\gamma$-ray energies, has revealed that the GRB afterglow emission is best described by a two component synchrotron shock model, which is highly suggestive of the standard forward/reverse shock interpretation \citep{laskar13,maselli14,panaitescu13,perley14}. In this picture the forward shock causes the emission at optical and X-ray frequencies after $~0.1$~days, and the radio emission after a few weeks, while the earlier emission at those frequencies is dominated by the reverse shock. Emission from the forward shock may also extend into the high-energy $\gamma$-ray regime, but this requires changes in the models of particle acceleration up to these very high energies \citep{ackermann14,kouveliotou13}.

The earliest VLA observations began $\sim0.68$ days post burst and yielded detections at 5.1 and 6.8 GHz \citep[][]{laskar13,perley14}. The 5~GHz VLA light curve also indicates a peak, but at $\sim2$~days, and the very well sampled light curve at this frequency taken with the Westerbork Synthesis Radio Telescope (WSRT) shows that this peak occurs at around $1.6$~days post-burst (van der Horst et al., in prep.). If the peak at 5 GHz is real then comparing this light curve with the AMI 15.7 GHz light curve of GRB 130427A shows the progression of the peak of the reverse shock occurring at later times with decreasing frequency, which is expected as a generic feature of all synchrotron models \citep[e.g.][]{vanderlaan66}. However, based on the VLA observations alone it is not possible to rule out interstellar scintillation as the cause of the 5 GHz peak since the source size at this time is small enough for scintillation effects to still be influencing the observed flux \citep{perley14}. 

\begin{figure}
\begin{center}
\includegraphics[width=0.5\textwidth]{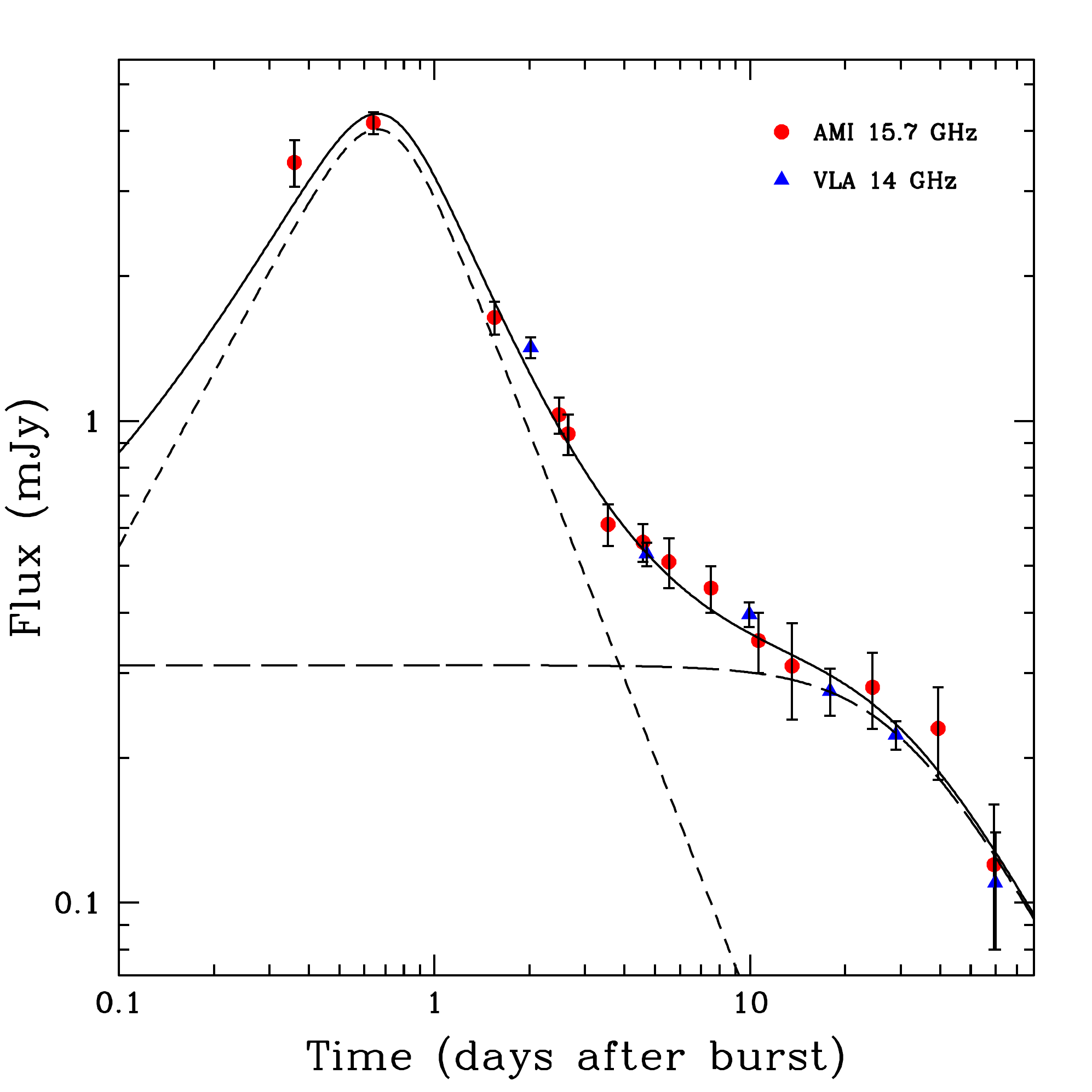}
\end{center}
\caption{The AMI 15.7 GHz and VLA 14 GHz light curve of GRB 130427A over plotted with the afterglow model derived by \citet{perley14} (solid line) showing the individual contributions from the reverse shock (short dashed line) and forward shock (long dashed line). The AMI peak at $\sim0.7$ days is one of the earliest radio peaks every observed from a GRB.}
\label{fig:2}
\end{figure}

The earliest VLA observations at 5.1 and 6.8 GHz, 0.68~days post-burst, suggest that the spectrum of GRB 130427A may have been affected by synchrotron self-absorption at this time. This is demonstrated by the steep spectral index $\beta \approx 2.4$ (for $S_{\nu} \propto \nu^{\beta}$) between these two frequencies. These VLA observations coincide with the 15.7 GHz flux peak seen in the second AMI observation at 0.64 days. A comparison between the VLA and AMI fluxes demonstrate that the spectrum was indeed increasing with frequency at that time between 5.1 and 15.7~GHz ($\beta \approx 1$), although not as steeply as between 5.1 and 6.8~GHz. This indicates that the peak of the spectral energy distribution of GRB 130427A was close to 15.7~GHz around $\sim0.7$ days post-burst, which is also confirmed by the peak observed in the 15.7~GHz light curve. The early radio peak in the AMI light curve is therefore caused by the synchrotron self-absorption turn-over frequency ($\nu_a$) moving through the observing bands rather than by the peak frequency \citep[$\nu_m$; for details regarding the GRB characteristic frequencies see][]{sari99}.

The double broken power-law fit to the AMI data in Figure~\ref{fig:1} shows a dramatic slow-down in the power-law decay of the light curve at 3.2 days post-burst. Such a change in the power-law index is atypical when compared to the power law breaks seen from GRBs at shorter wavelengths, which instead steepen due to the slowing of their collimated (jet) outflows \citep{sari99jet}. This flattening in the light curve decay suggests the presence of a second spectral component that is beginning to rise at late times in the radio band. This second component is interpreted by \citet{perley14} as the forward shock beginning to dominate over the reverse shock emission. In Figure~\ref{fig:2} we show a model light curve comprised of a reverse shock and a forward shock component. Note that this is not a formal fit to the 15.7~GHz light curve, which is in fact not possible given the large number of free parameters compared to the number of light curve data points, but is instead based on the fit results presented by \citet{perley14}, which assumes $\nu_a$ is causing the reverse shock peak. 

\citet{perley14} argue that the radio to X-ray broadband spectra show the jet from GRB 130427A is moving through a medium structured like a stellar wind that has an electron energy index of $p\simeq2.14$. \citet{perley14} also use the resulting light curve slopes to calculate that the reverse shock is Newtonian (often described as the thin-shell case) in which the shock does not become relativistic while crossing the shell behind the forward shock. In this scenario the Lorentz factor of the shock depends on the radius as $\Gamma\propto r^{-g}$ \citep{kobayashi00}, and therefore the light curve slopes depend on the free parameter $g$, which is specific to this model. The theoretical range of allowed $g$ values is  $1/2<g<3/2$ in the case of a stellar wind medium \citep[see][]{zou05}. Since we have shown based on the AMI and VLA observations between $0.64-0.68$ days that the peak of the 15.7~GHz light curve is caused by the passage of $\nu_a$, the theoretical pre-peak slope is $(25g+40)/(28g+14)$ and the post-peak slope is $-((15g+24)p+7g)/(28g+14)$ \citep{zou05}. In \citet{perley14} it is shown that the broadband light curves are best fit with $g\simeq3$, resulting in pre-peak and post-peak slopes of $1.2$ and $-1.7$, respectively (see Figure~\ref{fig:2}). We have adopted this value for $g$ even though it is outside of the theoretically allowed range \citep[as also pointed out by][]{laskar13,panaitescu13}. The maximum allowed value of $g\simeq3/2$ results in too steep light curve slopes of $1.4$ and $-2.0$. We note, however, that these theoretical light curve slopes have been derived assuming that we are observing straight into the jet. A viewing angle away from the jet axis but smaller than the jet opening angle could remedy this discrepancy. 

For the model curves shown in Figure~\ref{fig:2} it is clear that the forward shock is contributing to the total flux at all times, but starts to dominate the 15.7 GHz emission after 3 days. \citet{perley14} have shown that $\nu_a$ of the forward shock is below this observing band. The turn-over in the light curve, which \citet{perley14} calculated to be at $\sim30$ days, is therefore caused by $\nu_m$ passing through, when the forward shock post-peak light curve slope changes from $0$ to $-(3p-1)/4\simeq-1.4$ \citep{meszaros98}.

\section{Conclusions}

The AMI light curve of GRB 130427A agrees well with the forward/reverse shock interpretation suggested by many authors such as \citet{perley14}, \citet{panaitescu13}, and \citet{laskar13}. These early time AMI observations (within one day post-burst) have enabled us to not only obtain one of the earliest detections of a long GRB, but also capture the peak in the reverse shock emission at 15.7~GHz. This result has allowed us to further constrain the possible models of the early time afterglow from GRB 130427A by demonstrating that $\nu_a$, rather than $\nu_m$, is the cause of the radio light curve peak. This scenario will be further investigated by combining the AMI and VLA light curves with fine time sampling observations obtained with the WSRT in van der Horst et al. (in prep.).

The detection of the reverse shock radio peak in the AMI light curve of GRB 130427A clearly demonstrates the importance of rapid response radio follow-up programmes of GRBs. The AMI GRB follow-up programme \citep{staley13} is therefore crucial for exploring the early time radio signatures of GRBs and constraining the radio properties of these events within the first few hours post burst.

\section*{Acknowledgements}

We thank the staff of the Mullard Radio Astronomy Observatory for their invaluable assistance in the operation of AMI. Special thanks also goes to Natasha Hurley-Walker, Guy Pooley, and Yvette Perrott for their advice and help with this research and to the referee for their constructive response and suggestions. GEA, TDS, RPF acknowledge the support of the European Research Council Advanced Grant 267697 ``4 Pi Sky: Extreme Astrophysics with Revolutionary Radio Telescopes". AJvdH, RAMJW and AR acknowledge support from the European Research Council via Advanced Investigator Grant no. 247295. CR acknowledges the support of an STFC studentship. 

\label{lastpage}

\end{document}